\documentclass{iopconfser}
\usepackage{graphicx}
\usepackage{amsmath}

\begin{document}
	
\title{Negative piezoelectricity in quasi-two/one-dimensional ferroelectrics}

\author{Ning Ding\footnote{Corresponding author}, Shuai Dong\footnote{Corresponding author}}
\affil{Key Laboratory of Quantum Materials and Devices of Ministry of Education, School of Physics, Southeast University, Nanjing 21189, China}

\email{dingning@seu.edu.cn, sdong@seu.edu.cn}
	
\begin{abstract}
In recent years, the investigation of low-dimensional ferroelectrics has attracted great attention for their promising applications in nano devices. Piezoelectricity is one of the most core properties of ferroelectric materials, which plays the essential role in micro-electromechanical systems. Very recently, the anomalous negative piezoelectricity has been predicted/discovered in many quasi-two-dimensional layered ferroelectric materials. In this Topical Review, we will briefly introduce on the negative piezoelectricity in quasi-two/one-dimensional ferroelectrics, including its fundamental concept, typical materials, theoretical predictions, as well as experimental phenomena. The underlying physical mechanisms for negative piezoelectricity are divergent and varying from case by case, which can be categorized into four types. First, the soft van der Waals layer is responsible for the volume shrinking upon pressure while the electric dipoles is from non van der Waals layer. Second, the noncollinearity of local dipoles creates a ferrielectricity, which leads to orthogonal ferroelectric and antiferroelectric axes. Third, the electric dipoles come from interlayer/interchain couplings, which can be enhanced during the volume shrinking. Fourth, the special buckling structure contributes to local dipoles, which can be enhanced upon pressure. In real materials, more than one mechanism may work together. Finally, the future directions of negative piezoelectricity and their potential applications are outlooked.    
\end{abstract}
	
\section{Introduction}
Ferroelectrics are functional materials that have spontaneous macroscopic polarization and the polarization can be reversed by an external electric field. Due to the characteristics of non-volatility when power is turned off, and the fact that the phase transition temperature can generally reach room temperature, the research of ferroelectrics has received widespread attention~\cite{jiang2011-AM,naber2010-AM}. Over the past century, many advances have been made in ferroelectric physics and materials research. From the perspective of physical mechanism. Crystal symmetry group theory, phonon soft mode theory, and Landau phase transition theory support the framework of traditional ferroelectric physics and successfully describe a series of physical effects including ferroelectricity, piezoelectricity, and ferroelectric photovoltaic effects. From the perspective of materials. Various types of ferroelectrics have been reported including three-dimensional oxides/fluoride ferroelectrics, organic ferroelectrics and so on~\cite{ahn2004-science,waghmare2003-PRB,tanisaki1963-JPSP,tokunaga1966-PTP,lin2017-PRM,hill2000-JPCB}. From the view of application, ferroelectrics can be used in nonvolatile memories, field-effect transistors, ferroelectric photovoltaic devices, sensors and so on~\cite{scott2007-science,scott1989-science,Justin2013-JMCC,martin2016-NRM,wooten2000-JQE,muralt2000-JMM}.

The piezoelectric effect describes the linear electromechanical interaction between the mechanical and electrical states in materials without inversion symmetry, which has attracted great attention~\cite{fu2000-Nature,saito2004-Nature,wang2006-Science,ahart2008-Nature}. It is known that the ferroelectrics must be piezoelectrics, but piezoelectrics are not necessarily ferroelectrics. In other words, piezoelectricity is a property associated closely with ferroelectricity. There are many application based on piezoelectricity including the detection/generation of sonar waves, piezoelectric motors, various sensors and so on~\cite{setter2006-JAP,mao2014-AEM,scott2007-science}. The reverse piezoelectric effect illustrates the internal creation of a mechanical strain resulting from an external electric field. The positive and negative piezoelectric effects describe the change of polarization with deformation. Specifically, the positive piezoelectric effect widely exists in traditional perovskite ferroelectrics~\cite{thomann1990-AM}, namely the dipole of system decreases with the compressive stress. While the negative piezoelectric effect is less common and it exists in certain compounds~\cite{katsouras2016-NM,bernardini1997-PRB,shimada2006-JPAP,liu2017-PRL}. The piezoelectricity of ferroelectrics is interesting and require further exploration.
\subsection{Ferroelectricity in low-dimensional materials}
Since the graphene was exfoliated mechanically, two-dimensional materials have also been attracted great interest owing to the distinctive properties~\cite{novoselov2004-Science}. Thousands of two-dimensional materials exfoliated from the van der Waals (vdW) systems beyond graphene have been reported, which were first used to study the transport and photoelectric properties~\cite{chhowalla2015-CSR,lin2010-JPCL}. In addition, due to the reduction of dimension, the vdW two-dimensional materials can lose some symmetry compared with bulk structures, which may provide a potential way to design low-dimensional ferroelectrics. The rapid development of two-dimensional materials has also expanded the study of ferroelectrics to two-dimensional and even quasi-one-dimensional systems, which maintains the advantage of ferroelectrics in the post-Moore era~\cite{wang2023-NM,zhang2023-NRM,Wumenghao-wires-2018,guan-AEM-2020,qi2021-AM,wu2018-wires}.

According to the characteristics of two-dimensional materials close to the surface and easy to passivate, Wu \textit{et al} proposed to convert non-ferroelectric materials into ferroelectrics through surface functionalization~\cite{wu2012-JACS}. Specifically, the transition-metal molecular sandwich nanowires are functionalized, which break the spatial inversion symmetry and exhibit molecular reorientation in response to external electric field. In addition, graphene modified by hydroxyl is one of two-dimensional ferroelectrics, and the direction of polarization is controlled by that of hydroxyl molecular~\cite{wu2013-PRB,kan2013-APL}. Furthermore, more two-dimensional materials have been functionalized by various organic molecules, which is a general approach to design the two-dimensional ferroelectrics in vdW systems~\cite{jiang2015-ACR,sainsbury2012-JACS,voiry2015-NCh,zhou2014-RA}.

In addition to using the functional molecules modification to design two-dimensional ferroelectric, researchers have been committed to the search and design of intrinsic two-dimensional ferroelectric materials. Shirodkar \textit{et al} has observed the unstable phonon vibration modes in monolayer MoS$_2$, which induces the ferroelectric phase of $d1T$-MoS$_2$~\cite{shirodkar2014-PRL}. There are several intrinsic two-dimensional ferroelectric materials confirmed by experiment. In 2015, Belianinov \textit{et al} found the layered materials CuInP$_2$S$_6$ holds the room-temperature ferroelectricity~\cite{belianinov2015-NL}, and the effect of nanocrystal thickness on ferroelectric properties was studied~\cite{chyasnavichyus2016-APL}. The following experiment has been shown that the ferroelectricity also maintains even the thickness of CuInP$_2$S$_6$ films is reduced to $4$ nm~\cite{Liu-Nat-Com-2016}. And the unusual ferroelectric property for the uniaxial quadruple potential well has been reported~\cite{brehm2020-NM}. Besides, Chang \textit{et al} observed the in-plane ferroelectric polarization in monolayer SnTe~\cite{chang2016-Science}, and the two-dimensional ferroelectricity has also been reported in other group IV monochalcogenides including GeS, GeSe, SnS and SnSe monolayers~\cite{fei2016-PRL,song2021-PRB}. In 2016, Ding \textit{et al} predicted the out-of-plane polarization in monolayer In$_2$Se$_3$ and other III$_2$-VI$_3$ materials~\cite{ding2017-NC}. Soon after, two-dimensional ferroelectricity in monolayer In$_2$Se$_3$ was confirmed by experiment~\cite{zhou2017-NL,xue2018-ACSNano}. Recently, the single element bismuth monolayer was synthesized and confirmed as a two-dimensional ferroelectric in experiment~\cite{gou2023-Nature}. Its structure is similar to the buckled black phosphorus and the origin of ferroelectricity is a combination of charge transfer and conventional atomic distortion~\cite{Xiao2018-AFM}. Furthermore, many intrinsic two-dimensional and quasi-one-dimensional ferroelectrics have been predicted theoretically~\cite{bruyer2016-PRB,xu2017-Nanoscale,zhangsh2018-Nanoscale,zhang2019-JACS,sun2022-PRM,song2024-JMCC,yang2021-ACSAMI,wang2021-PRM}.

The concept of sliding ferroelectricity was proposed by Wu \textit{et al} in 2017~\cite{li2017-ACSNano}. It indicates the unique stacking modes among layers can break the spatial inversion symmetry and induce the out-of-plane polarization, which can be reversed by interlayer sliding. This discovery expands the study of two-dimensional ferroelectrics from few specific materials to most of the currently known two-dimensional materials. Soon after, the sliding ferroelectricity in bilayer BN verified by experiment~\cite{yasuda2021-Science,vizner2021-Science}. More and more interlayer sliding ferroelectrics have been predicted and confirmed~\cite{yang2018-JPCL,liu2019-Nanoscale,Fei-Nature-2018,sharma2019-SA,xiao2020-NP,wang2022-NN,wan2022-PRL,sui2023-NC,miao2022-NM,liu2020-PRL,zhang2021-PRB, ma2021-npjCM, ding2021-PRM, chen2024-NL, meng2022-NC, liu2023-npjCM,yang2023-PRL, liang2021-npjCM, xiao2022-npjCM, zhou2022-npjMA,wu2021-PNAS}. In fact, the layer group of possible bilayer stacking ferroelectrics and the rules about the creation and annihilation of symmetry operations can be listed by performing group theory analysis~\cite{ji2023-PRL}.

\subsection{Negative piezoelectricity}
\begin{figure}
	\centering
	\includegraphics[width=\textwidth]{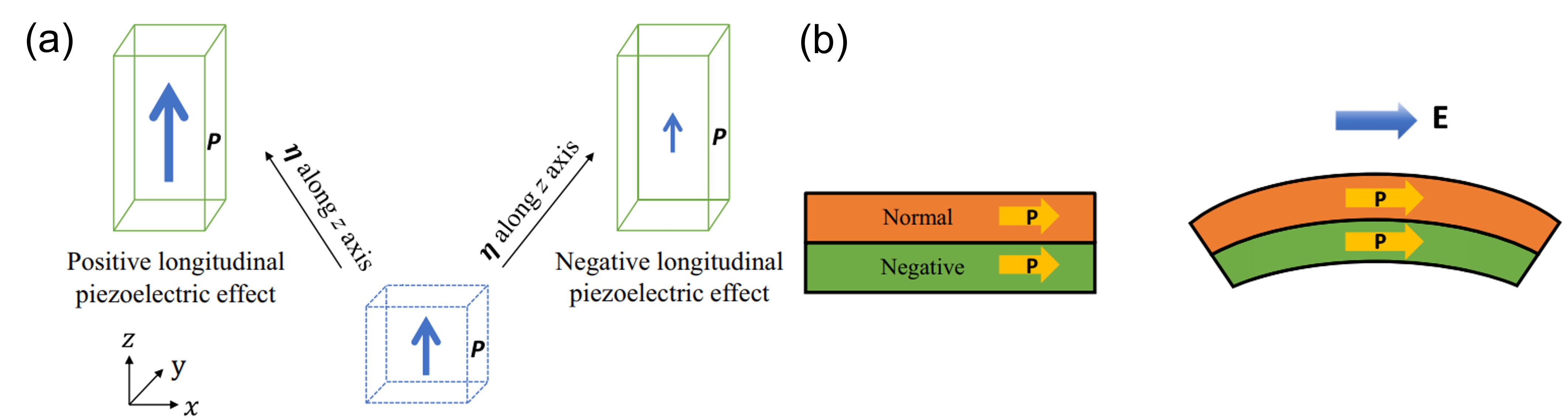}
	\caption{(a) Draft of positive and negative piezoelectricity. Assuming the $z$-axis is the direction of exterial strain. (b) Draft of heterostructure constructed by positive and negative piezoelectric layers. Reprint figure with permission from~\cite{wang2023-PRB}.}
	\label{fig-negative}
\end{figure}
Research on piezoelectricity has a history of one hundred years and materials expands from conventional bulk materials to low-dimensional systems. Piezoelectric property differs slightly in these previous two kinds of materials. First, the traditional bulk materials including perovskite oxide and fluoride are mainly connected through ionic or covalent bonds, which are “hard”, while the layers/chains in most low-dimensional systems are bonded by the weak van der Waals interaction, which are “soft”. The difference of structure can strongly affect the change in polarization due to structural relaxation in response to strain, namely internal-strain term. This may result in the ubiquity of negative piezoelectricity in layered ferroelectric materials \cite{qi2021-PRL}. Second, piezoelectricity in low-dimensional materials exhibits strong anisotropy due to the reduced dimensionality \cite{Cui2024-npj-2D,zhao2021-JMCC,Fei2015-APL}. Third, morphotropic phase boundary (MPB) has been widely used to explain the giant piezoelectricity in conventional bulk materials, while MPB in low-dimensional systems has received very little attention perhaps due to challenges in achieving phase coexistence and possible polarization rotation \cite{song2021-PRB,Deng2024-2D-Materials,wang2024-NL}.  

Piezoelectricity is usually estimated by the piezoelectric coefficient that describes how the polarization changes in response to a stress or strain. This derives two physical quantities, namely piezoelectric stress coefficient $d_{ij}$ and the piezoelectric strain coefficient ($e_{ij}$). Specifically, $d_{ij}$ describes the induced polarization in direction $i$ ($\Delta P_i$) with an applied stress ($\sigma_j$), which can be gauged by $\Delta P$=$d_{ij}\sigma_j$. And $e_{ij}$ links $\Delta P_i$ with strain ($\eta_j$) given by $\Delta P_i$=$e_{ij}$$\eta_j$. Both $d_{ij}$ and $e_{ij}$ are third-rank tensors, and they can be linked via elastic compliances by $d_{ij}$=$S_{jk}$$e_{ik}$~\cite{liu2017-PRL}. The positive (negative) longitudinal piezoelectric coefficients ($e_{33}$ and $d_{33}$, assuming the $z$-axis is the direction of polar axis) represent positive (negative) piezoelectric effect. As depicted in the Fig.~\ref{fig-negative}(a), a uniaxial tensile strain increases (decreases) the polarization for positive (negative) piezoelectric effect~\cite{wang2023-PRB}. From perspective of materials, the positive piezoelectrics are common. There is an exception, namely the ferroelectric polymerpoly (vinylidene fluoride) (PVDF), which exhibits negative piezoelectric effect~\cite{katsouras2016-NM,bystrov2013-JMM}. From perspective of application, negative piezoelectrics are unique and promising in the devices used for electromechanical system. As depicted in Fig.~\ref{fig-negative}(b), the heterostructure constructed by positive and negative piezoelectrics can achieve better ductility.

Recently, more and more quasi-two/one-dimensional ferroelectrics have been reported, and the unique negative piezoelectric mechanisms have also been discovered~\cite{you2019-SA,lin2019-PRL,ding2021-PRM,wang2023-PRB,zhong2023-PRL}. In this brief review, the negative piezoelectricity in quasi-two/one-dimensional ferroelectrics reported recently will be introduced and categorized. 
	
\section{Negative piezoelectricity in low dimensional ferroelectrics}
In the following, we will review several works about negative piezoelectricity in low dimensional ferroelectrics with different physical mechanisms. The origin of negative piezoelectricity is closely related to that of ferroelectricity. The following examples include vdW layered ferroelectric CuInP$_2$S$_6$, noncollinear, sliding, and elemental ferroelectrics.   
	
\subsection{Gaint negative piezoelectricity in CuInP$_2$S$_6$}
\subsubsection{Ferroelectricity of CuInP$_2$S$_6$}
\begin{figure}
	\centering
	\includegraphics[width=0.7\textwidth]{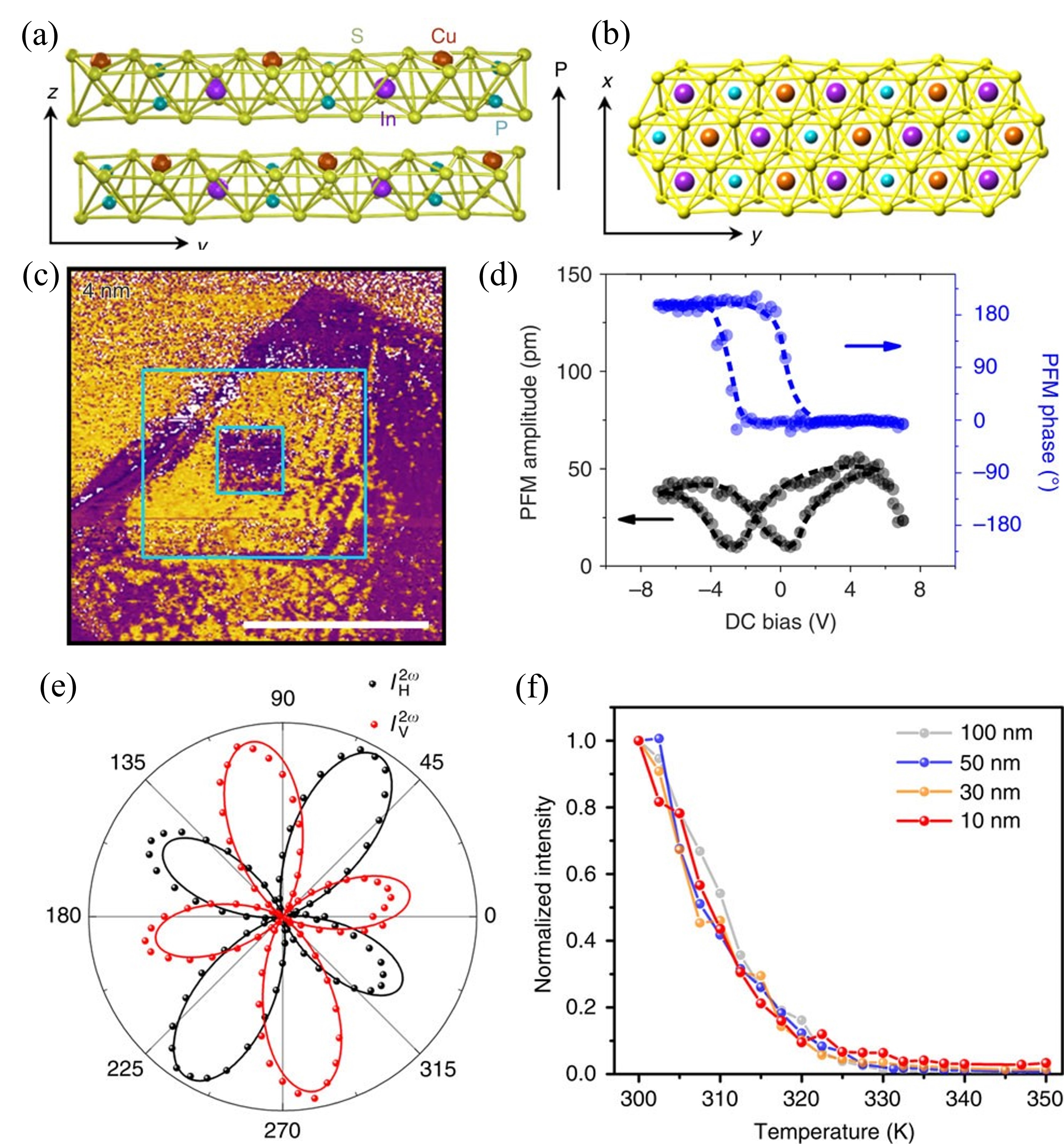}
	\caption{(a) Side view and (b) top view for the structure of bulk CIPS. The arrow indicates the polarization direction. (b) The PFM phase image for $4$ nm. (d) The phase (bule) and amplitude (black) hysteresis loop from PFM for $4$ nm. (e) SHG intensity in horizontal (H) and vertical (V) directions for a $100$ nm. (f) Temperature dependence of SHG intensity with different thickness~\cite{Liu-Nat-Com-2016}.}
	\label{fig-CIPS-FE}
\end{figure}
The ferroelectricity of CuInP$_2$S$_6$ (CIPS) has been confirmed experimentally and theoretically, which has attracted intensive attention recently~\cite{belianinov2015-NL,chyasnavichyus2016-APL,Liu-Nat-Com-2016,brehm2020-NM}. CIPS is one of vdW layered materials that remains room-temperature ferroelectricity when the thickness is reduced to $\sim 4$ nm, whose structure can be depicted in Fig.~\ref{fig-CIPS-FE}(a,b). Monolayer of CIPS contains of a sulfur framework stuffed by Cu, In ions and P-P pairs. The bulk system is made up of monolayers assembled by -A-B-A-B- mode along the $z$-axis, which can be coupled by vdW interaction. The site exchange between Cu and P-P pair causes two different layers, A and B. The displacement of Cu and other cations converts the space group from $C2/c$ (non-polar) to $Cc$ (polar) with spontaneous out-of-plane polarization.
	
An important criterion of ferroelectric is whether the polarization can be switched by an electric field. Piezoresponse force microscopy (PFM) measurement has been used as depicted in Fig.~\ref{fig-CIPS-FE}(c,d). The butterfly loop of the PFM amplitude signal and the obvious $180^\circ$ flopping of the phase signal indicates the out-of-plane ferroelectric polarization in CIPS even the thickness reduced to $\sim 4$ nm. In addition, second-harmonic generation (SHG) measurement has been used to probe the phase transition from the ferroelectric to paraelectric state in experiment, as shown in Fig.~\ref{fig-CIPS-FE}(e,f). These experimental results verify the ferroelectricity in few layer CIPS.
	
\subsubsection{Negative piezoelectricity of CIPS confirmed by experiment}
\begin{figure}
	\centering
	\includegraphics[width=\textwidth]{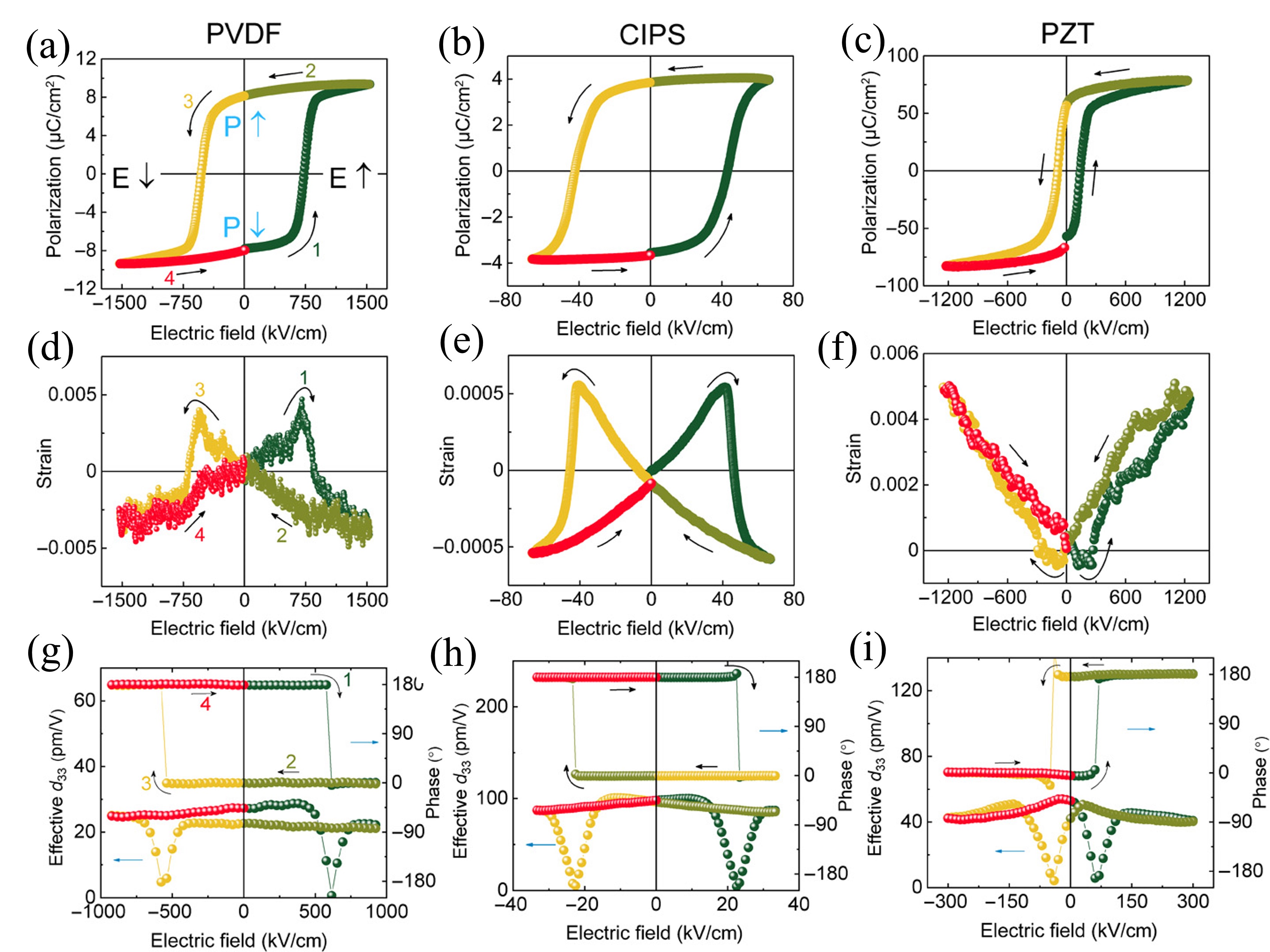}
	\caption{(a-c) Polarization-electric field curves for PVDF (a), CIPS (b),and PZT (c). (d-f) Corresponding strain-electric loops for PVDF (d), CIPS (e), and PZT (f). (g-i) $d_{33}$ of PVDF (g), CIPS (h), and PZT (i). Reprint figure with permission from~\cite{you2019-SA}.}
	\label{fig-CIPS-S-E}
\end{figure}

Piezoelectricity is a key property associated with ferroelectricity, and it is natural to investigate the piezoelectricity of CIPS~\cite{you2019-SA,qi2021-PRL}. Besides, the anomalous polarization enhancement in CIPS under pressure has been reported by the previous experiment \cite{Yao2023-NC}. The converse piezoelectric effect of CIPS is measured by experiment. For comparison, the results of two other prototypical ferroelectric materials are exhibited, $i. e.$ PVDF and Pb(Zr$_x$Ti$_{(1-x)}$)(PZT) thin films. There two materials are known examples of negative piezoelectric and positive piezoelectric respectively. The polarization, strain and longitudinal piezoelectric coefficient $d_{33}$ as the function of electric field are synchronously recorded as depicted in the Fig.~\ref{fig-CIPS-S-E}.
	
In the PVDF system, the polarization is downward at the beginning. When the electric field increases from zero to the maximum pointing upward, polarization has been switched at the coercive field ($E_{\rm c}$). Simultaneously, the strain increases from zero to the peak at the $E_{\rm c}$, and flips the symbol after the polarization completely switched. This phenomenon indicates the lattice expands (contracts) if the electric field is opposite (same) to the direction of polarization. This electromechanical response of PVDF is consistent with negative piezoelectric effect. On the contrary, the strain-electric field (S-E) loop for PZT is opposite to PVDF, characterized as the positive piezoelectric effect. While for CIPS, its electromechanical response seems to be the similar with PVDF, indicating it is a negative piezoelectric material.
	
To double-check this discovery, the piezoelectric coefficients of $d_{33}$ are measured by experiment. PVDF and CIPS have the same behavior, which agree with the negative piezoelectric effect, while PZT exhibits the positive piezoelectric effect. The effective $d_{33}$ in zero-field case of PVDF, CIPS, and PZT are about $-25$, $-95$, and $48$ pm/V, respectively. The large piezoelectric coefficient $d_{33}$ of CIPS indicates the giant negative piezoelectric effect, whose physical mechanism should be revealed.
	
\subsubsection{Phenomenological theory of the negative piezoelectricity in CIPS}
The negative piezoelectric effect can be revealed by phenomenological theory. A ferroelectric crystal changes from the paraelectric phase to the polar state, which can induces the longitudinal strain named as $S_{33}$. It can be estimated by the Taylor expansion of the electric distortion named as $D_3$ as shown in the following equation~\cite{lines2001principles}:
\begin{equation}
	S_{33}=Q_{33}D^2_3=Q_{33}(\vec{P_s}+\vec{P_i})^2=Q_{33}(\vec{P_s}+\epsilon_{33}\vec{E_3})^2=Q_{33}\vec{P_s}^2+2Q_{33}\epsilon{33}\vec{P_s}\vec{E_3}+Q_{33}\epsilon_{33}^2\vec{E_3}^2
\end{equation}
where $Q_{33}$ is the coefficient of longitudinal electrostriction, $\epsilon_{33}$ is the dielectric permittivity, $\vec{P_s}$ and $\vec{P_i}$ are the spontaneous and induced polarization. The spontaneous strain induced from the paraelectric-to-ferroelectric phase transition can be described in the first term. The main electromechanical coupling component can be represented by the second term, in other word, piezoelectric coefficient of $d_{33}$ can be defined as $2Q_{33}\epsilon_{33}P_s$. Quadratic electrostriction effect can be shown in the last part~\cite{li2014-APR}. The negative piezoelectricity in ferroelectric systems is the spontaneous negative electrostriction, which can be shown as the expandation of lattice constant in the polar axis when the drop of polarization.
	
\subsubsection{Revealing negative piezoelectricity from dimensionality}
\begin{figure}
	\centering
	\includegraphics[width=\textwidth]{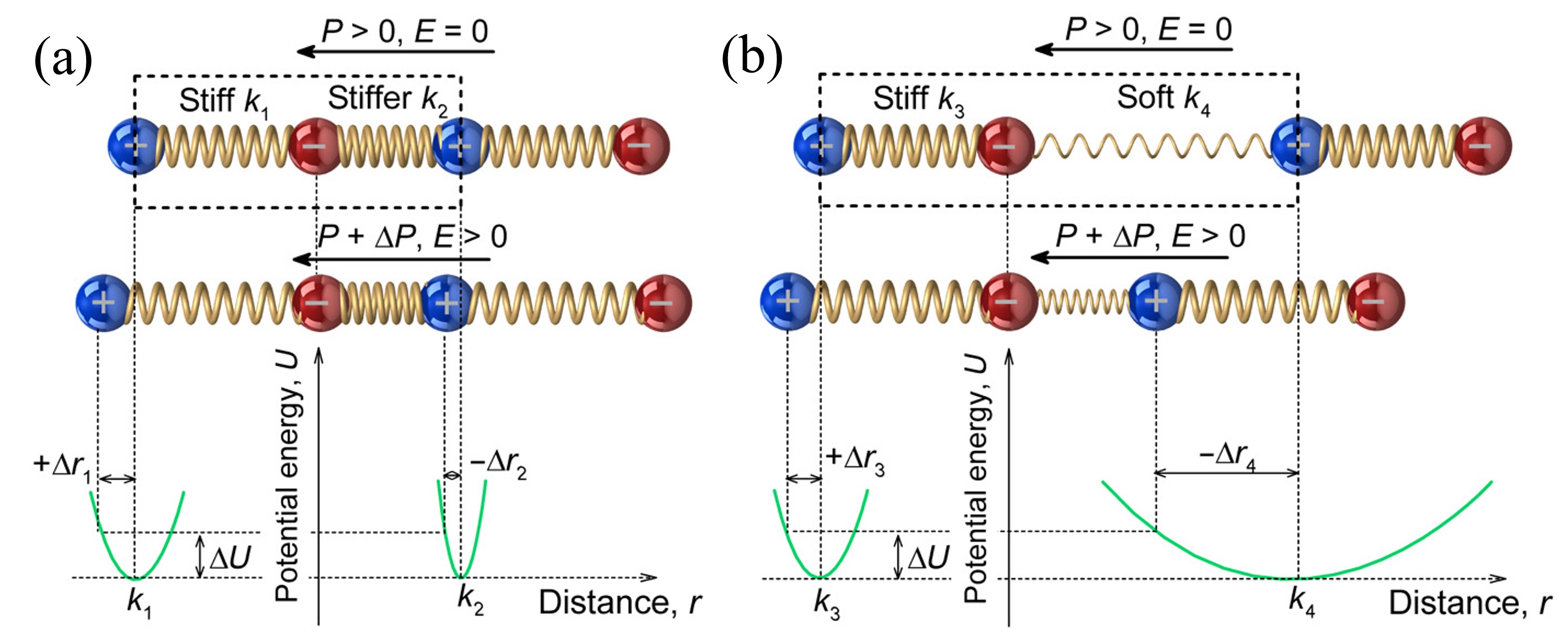}
	\caption{Simplified rigid ion model for piezoelectrics.(a) Positive piezoelectrics with continuous lattice. (b) Negative piezoelectrics with discontinuous lattice. The dashed box indicates the unit cell. The black arrows indicate the directions of polarization and electric field. The pair potential energy profiles of the relevant chemical bonds are shown by the lower parts. Reprint figure with permission from~\cite{you2019-SA}.}
	\label{fig-CIPS-elastic}
\end{figure}

Different from the perovskite oxide three-dimensional ferroelectrics, the dipole in CIPS exists in the isolated layers, which can be stacked together via vdW interaction to form the discontinuous (broken) lattice. The ions and bonds have been simplified to the ball-and-spring model as depicted in Fig.~\ref{fig-CIPS-elastic}. In PZT, the electric field and polarization are assumed to be along the same direction, the positive (negative) ions move along (against) the electric field. While due to the bond anharmonicity of spontaneous displacement, namely, the compression is harder than the expansion. This results in $\Delta r_1>\Delta r_2$, and the lattice parameter expands with the enhancement of polarization. Therefore, it is the positive piezoelectricity. However, in CIPS, the isolated dipoles are held together via the vdW interaction, which is very soft. So, the elastic compliance of $k_4$ is very small, resulting in $\Delta r_4>\Delta r_3$. This means the lattice parameter shrinks with the enhancement of polarization in CIPS, namely the negative piezoelectricity.
	
\subsubsection{Microscopic origin of negative piezoelectricity in CIPS from computational modeling}
\begin{figure}
	\centering
	\includegraphics[width=\textwidth]{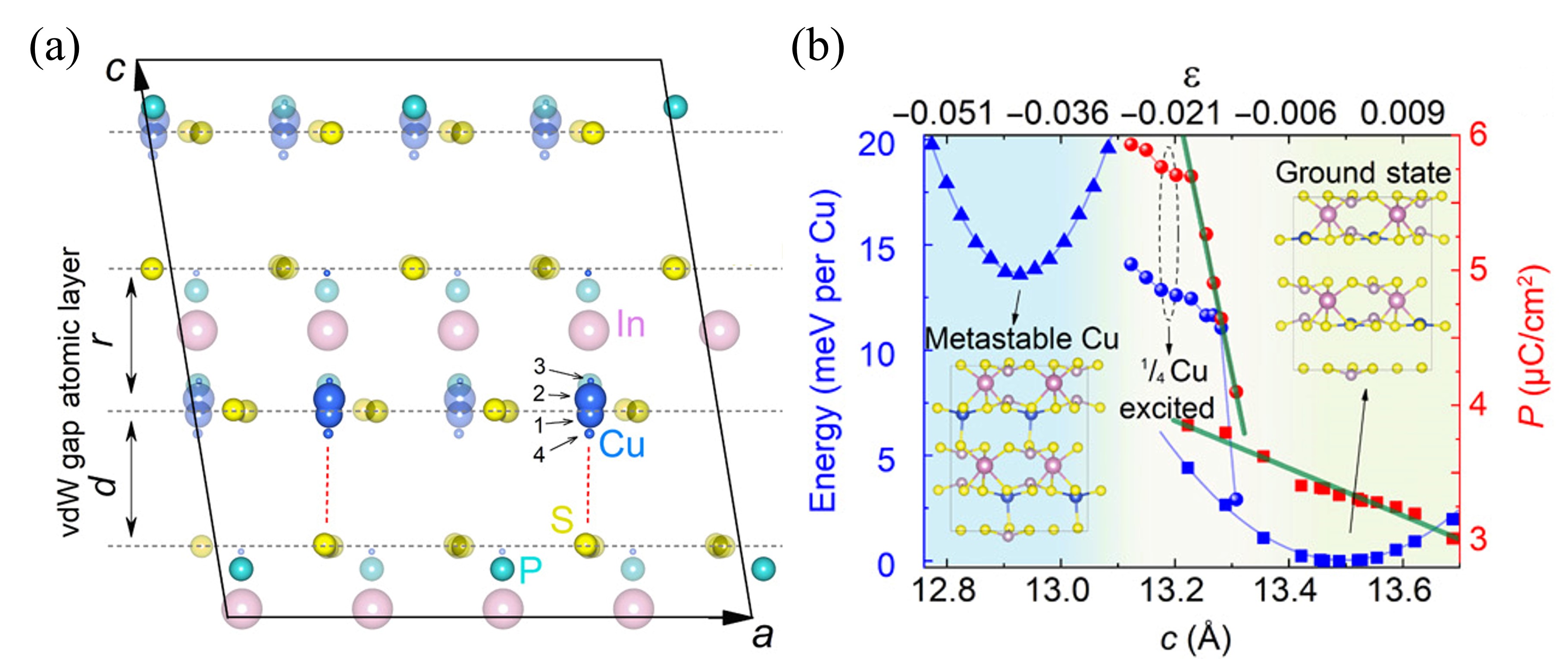}
	\caption{(a) Refined structure of CIPS in unit cell viewed from the $b$ axis. The occupation can be indicated by the Cu atoms' size. (b) The free energy $E$ (left vertical coordinate) and polarization $P$ (right vertical coordinate) changing with lattice constant $c$ in different cases: triangle, metastable state occupied by all Cu ions; circle, metastable state occupied by $\frac{1}{4}$ Cu ions; square, ground-state. Insets: the relevant structures. Reprint figure with permission from~\cite{you2019-SA}.}
	\label{fig-CIPS-DFT}
\end{figure}
The single-crystal X-ray diffraction and first-principles calculations have been performed to reveal the giant negative piezoelectric effect from a microscopic perspective. There are four Cu sites in one of the polarization state, as shown in the Fig.~\ref{fig-CIPS-DFT}(a). In addition, experiment result suggests there is a possible metastable Cu site in the vdW gap. And the occupancy of interlayer site is estimated to $0.08$.
	
For the first-principles calculation, the piezoelectric coefficient $d_{33}$ of the ground state is $-18$ pC/N, which is far from the experimental value ($d_{33}\approx-95$ pC/N). The giant piezoelectricity in CIPS can't be explained just by the ground state. Then the partial occupancy has been taken into consideration, as depicted in Fig.~\ref{fig-CIPS-DFT}(b). In addition, taking the occupancy ($8\%$) observed by experiment into consideration, the estimated polarization is $\sim 4.15 $ $\mu$C/cm$^2$, which is consistent with the experimental value. Furthermore, the approximate $d_{33}$ is about $-110$ pC/N. The first-principles calculations confirm the shrinking of the distance of vdW layers $d$ results in most of the change in $c$ lattice constant, while the CIPS layer thickness $r$ barely changes, which is consistent with revealing the negative piezoelectricity from dimensionality.
	
There are two important conclusions from the first-principles calculations. First, the shrinking of the vdW layers plays an important role in the negative piezoelectricity of CIPS. Second, the giant magnitude of piezoelectric coefficient originates from the large displacive instability of the Cu atoms.   
	
\subsection{Negative piezoelectricity in noncollinear ferrielectrics}
\subsubsection{The noncollinear ferrielectricity in monolayer $M$O$_2$$X_2$}
\begin{figure}
	\centering
	\includegraphics[width=1\textwidth]{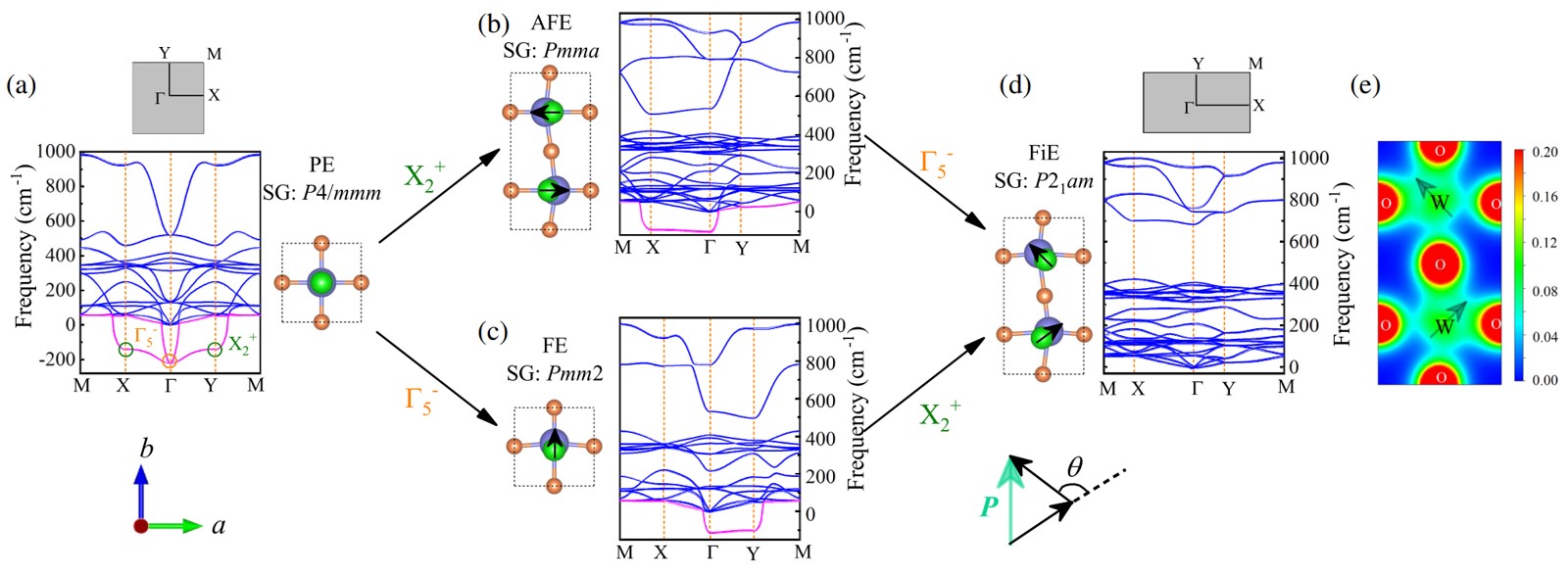}
	\caption{Phonon spectra and structures for different phases of monolayer $M$O$_2$$X_2$. Brillouin zones are represented by the grey squares or rectangles. (a) The paraelectric phase. The imaginary-frequency modes ($X_2^{+}$ and $\Gamma_5^{-}$) are indicated. (b) The unstable antiferroelectric state accompanying the distortion mode named $X_2^{+}$. (c) The unstable ferroelectric state accompanying the distortion mode named $\Gamma_5^{-}$. (d) The final ferrielectric ground state. The angles marked by $\theta$ between nearest neighbor dipoles (black arrows) align the $b$ axis are $\sim 129^\circ$. (e) Partial charge density diagram of valence electrons. Reprint figure with permission from~\cite{lin2019-PRL}.}
	\label{fig-PRL-phonon}
\end{figure}
Magnetism and polarity are homotopic according to the Landau theory, which should display the one-to-one correspondence among many physical properties. For example, ferromagnetic $\it{vs}$ ferroelectric states, antiferromagnetic $\it{vs}$ antiferroelectric states. The spin noncollinearity has been widely studied for magnets, while the noncollinear dipole orders are very rare except a few ferrielectrics, such as three-dimensional materials of BaFe$_2$Se$_3$~\cite{dong2014-PRL}, Pb$_2$MnWO$_6$~\cite{orlandi2014-IC}, and monolayer $M$O$_2$$X_2$ ($M$: group-VI transition metal; $X$: halogen)~\cite{lin2019-PRL}. The noncollinear ferrielectricity associated with negative piezoelectricity in monolayer $M$O$_2$$X_2$ will be reviewed in the following.
	
The paraelectric structure of $M$O$_2$$X_2$ indicates all $M$ ions are located at the center of the O$_4$$X_2$ octahedra with the space group of $P4/mmm$ (shown in Fig.~\ref{fig-PRL-phonon}(a)). The phonon spectrum indicates the paraelectric structure is unstable with two imaginary-frequency modes, which can result in the spontaneous structural distortions. As depicted in Fig.~\ref{fig-PRL-phonon}(b,c), the unstable mode named $X_2^{+}$ at $X$ (and $Y$) induces antiferroelectric distortions and another unstable mode named $\Gamma_5^{-}$ at $\Gamma$ point results in ferroelectric distortions. 
	
While, only one of phonon modes can't construct a stable phase. The cooperation of these two modes can lead to a stable phase with the net ferrielectricity [Fig.~\ref{fig-PRL-phonon}(d)], whose behavior is similar with the exchange frustration in magnets. The dipole moments at local sites are noncollinear, and the canting angle of WO$_2$Cl$_2$ (MoO$_2$Br$_2$) is about $129^\circ$ ($128^\circ$) as depicted in the bottom of Fig.~\ref{fig-PRL-phonon}. The directions of ferroelectric and antiferroelectric ordering are orthogonal, along the $b$ and $a$ axes. The polarizations in the $b$-axis of monolayer MoO$_2$Br$_2$ and WO$_2$Cl$_2$ are estimated to $32.09$ and $26.36$ $\mu$C/cm$^2$ respectively when the distance between adjacent layers as the thickness of a monolayer.
	
The driving force of polar distortions of these transitional metal oxyhalides should be the $d^0$ rule from $M^{6+}$, which means strong coordination bonds can be formed between the empty $d$ orbitals of $M$ ions and $2p$ orbitals of $O$ ions. As shown in Fig.~\ref{fig-PRL-phonon}(e), there is a few valence electrons transfers from O$^{2-}$ to $M^{6+}$ and this induces the ferrielectric distortions.
	
\subsubsection{Negative piezoelectricity induced by the noncollinear ferrielectricity in monolayer $M$O$_2$$X_2$}
\begin{figure}
	\centering
	\includegraphics[width=1\textwidth]{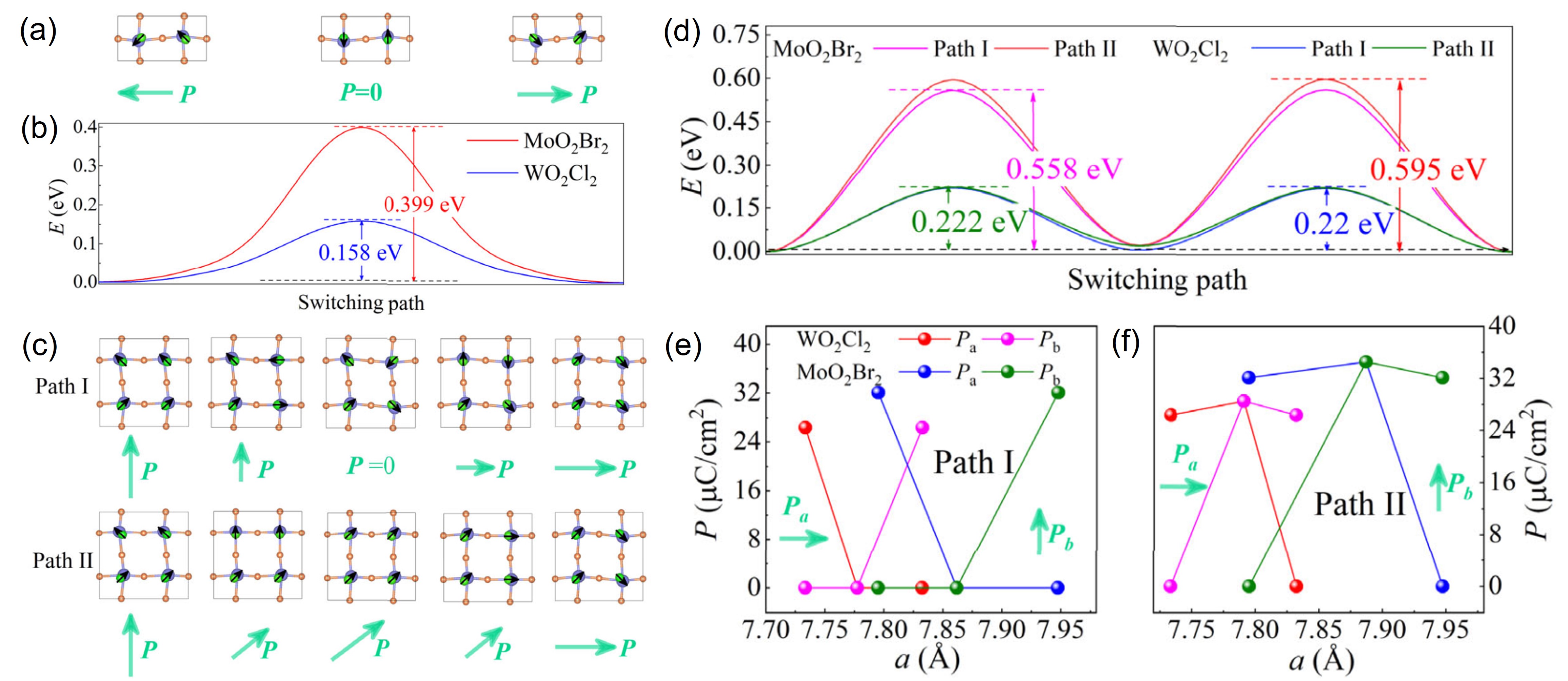}
	\caption{(a) The $180^\circ$ switching path of the net polarization: ferrielectric-antiferroelectric-ferrielectric. (b) The corresponding energy barriers of the $180^\circ$ switching of polarization. (c) Two assuming paths (labeled as Path I and Path II) of the $90^\circ$ switching of polarization by interchanging the $a$ and $b$ axes. (d) The corresponding energy barriers of the $90^\circ$ switching. (e,f) The change of local dipoles along the $a$ ($P_a$) and $b$ ($P_b$) directions as the function of lattice constant $a$. The general trend is that $P_a$ reduces with the enlarge of $a$, namely the negative piezoelectricity. (e) Path I. (f) Path II. Reprint figure with permission from~\cite{lin2019-PRL}.}
	\label{fig-PRL-switching}
\end{figure}
Different from the plain ferroelectrics, the $180^\circ$ flipping of the net polarization doesn't need the $180^\circ$ switching of every local dipole moments. On the contrary, the collaborative $\sim 50\%$ rotations of the dipole moments on every local sites can also work [Fig.~\ref{fig-PRL-switching}(a)]. In addition, the energy barriers induced by the nonpolar transition states for MoO$_2$Br$_2$ and WO$_2$Cl$_2$ monolayers are $0.399$ and $0.158$ eV, respectively [Fig.~\ref{fig-PRL-switching}(b)].
	
Furthermore, the $90^\circ$ switching of polarization also works, which can be achieved by exchanging the $a$ and $b$ axes. There are two assuming two-step paths can be shown in Fig.~\ref{fig-PRL-switching}(c). And the energy barriers can be depicted in Fig.~\ref{fig-PRL-switching}(d), which induces a slightly lower energy barrier in path II. This unique $90^\circ$ switching of polarization can further result in the negative piezoelectricity. The reason can be explained as the following. The lattice constants $a$ and $b$ in the orthorhombic ferrielectric phase are unequal, i.e. $a>b$, and the net polarization is along the $b$-axis. While the component of local dipoles along the $a$-axis is larger than that along the $b$-axis. Thus, when the net residual polarization undergoes the $90^\circ$ switching process from the $b$-axis to $a$-axis, the lattice constant along the polarization increases, which means the $P_a$ decreases with increasing $a$ [Fig.~\ref{fig-PRL-switching}(e,f)], and the anomalous behavior namely the negative piezoelectricity appears. In other words, the negative piezoelectric effect can be induced by the $90^\circ$ switching of polarization. However, the specific magnitude of negative piezoelectric coefficient has not been proven by calculation or experiment, which needs further study.

\subsection{Negative piezoelectricity in sliding ferroelectrics}
\subsubsection{Ferroelectricity in interlayer sliding ferroelectrics}
\begin{figure}
	\centering
	\includegraphics[width=\textwidth]{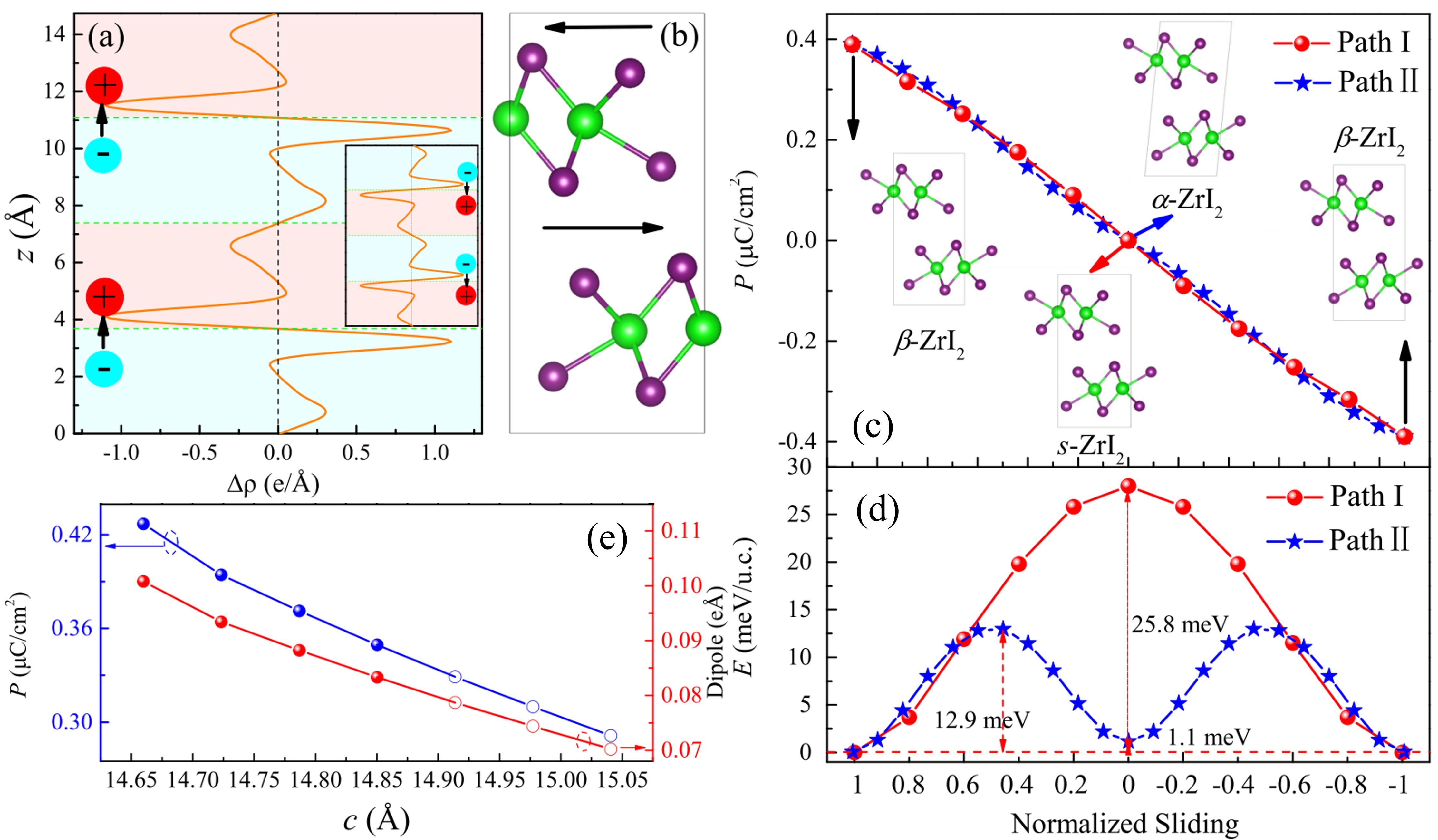}
	\caption{(a) The planar-averaged differential charge density ($\Delta\rho$) projecting along the $c$-axis, with parent phase (space group: $Pnma$) as the reference. The case of opposite polarization can be shown in the inset. (b) The corresponding structure of (a). (c) Two possible ferroelectric switching paths in $\beta$-ZrI$_2$. (d) The corresponding energy barriers. (e) The polarization and dipole moment of $\beta$-ZrI$_2$ as the function of lattice constant $c$. Reprint figure with permission from~\cite{ding2021-PRM}.}
	\label{fig-ZrI2}
\end{figure}
Interlayer sliding ferroelectricity is widely exists in two-dimensional vdW materials, which describes the unique phenomenon that the induced ferroelectric polarization perpendicular to the interlayer sliding direction. Sliding ferroelectrics have been received extensive experimental and theoretical attention. The origin of sliding ferroelectricity and negative piezoelectricity will be reviewed. 

Taking $\beta$-ZrI$_2$ bulk as an example of sliding ferroelectric~\cite{ding2021-PRM}, whose polarization is calculated as $0.39$ $\mu$C/cm$^2$ by first-principles calculations, whose polarizatio direction is along the $c$-axis. The planar-averaged differential charge density between the ferroelectric and paraelectric (with space group of $Pnma$) has been calculated and can be shown in Fig.~\ref{fig-ZrI2}(a). It is obvious that there is charge transfer between upper and lower Zr-I bonds in each layer, which should be induced by the special stacking way with symmetry breaking. In other words, the inequivalent of chemical environment makes upper Zr-I bonds more charge positive but the lower side more negative and it naturally induces a electric dipole moment in each ZrI$_2$ layer. Different from the ZrI$_2$ bulk, the ZrI$_2$ bilayer owns both in-plane and out-of-plane polarization. While the in-plane polarization can be offset between nearest-neighbor bilayer in bulk [Fig.~\ref{fig-ZrI2}(b)].

The direction of polarization can be switched by interlayer sliding. There are two ferroelectric switching paths named as path I and path II as depicted in Fig.~\ref{fig-ZrI2}(b). Path I is only related to the interlayer sliding among vdW layers. Path II involves not only interlayer sliding but also changes of crystallographic axis angle. Both paths can achieve the ferroelectric switching, but have different energy barriers as depicted Fig.~\ref{fig-ZrI2}(d).

\subsubsection{Negative piezoelectricity in sliding ferroelectrics}
The polarization can be regulated by the uniaxial stress (changing the lattice constant) align the $c$-axis as depicted in Fig.~\ref{fig-ZrI2}(e). It is obvious that the polarization increase as the lattice constant decreases, which indicates $\beta$-ZrI$_2$ has the negative piezoelectric effect. Furthermore, the piezoelectric coefficient $d_{ij}$ has been calculated as
\begin{equation}
	d_{ij}= {\textstyle \sum_{k=1}^{6}}e_{ik}S_{kj} 
\end{equation}
where $S$=$C^{-1}$ is the elastic compliance coefficients, and $C$ is the elastic tensor. The results show $d_{33}$ is $-1.445$ pC/N, which further confirm the negative piezoelectric by first-principles calculations.

Ther are two contributions to the enhancement of polarization accompanying the compression of lattice constant $c$. First term is the decreased volume due to the easily compressible soft vdW layers, which is similar to CIPS~\cite{you2019-SA}. Second term is the increased dipole moment of each unit cell, which is owing to the unique origin of interlayer sliding ferroelectricity. Specifically, the closer adjacent ZrI$_2$ layers, the larger interlayer sliding, and the greater vertical polarization. For example, in $\beta$-ZrI$_2$, the contributions of these two parts for negative piezoelectricity are $\sim 5\%$ and $\sim 95\%$, respectively. In addition, recent reports indicate that the negative piezoelectric effect widely exists in sliding ferroelectrics, including bilayer BN, bilayer WTe$_2$, 2D Tellurene \cite{Jiang2022-PRB,yang2018-JPCL,liu2019-Nanoscale,Sachdeva2024-JPCM} and quasi-one-dimensional sliding ferroelectric NbI$_4$ \cite{ding2024-PRB}.

\subsection{Quasi-one-dimensional sliding ferroelectricity and negative piezoelectric effect}
\begin{figure}
	\centering
	\includegraphics[width=\textwidth]{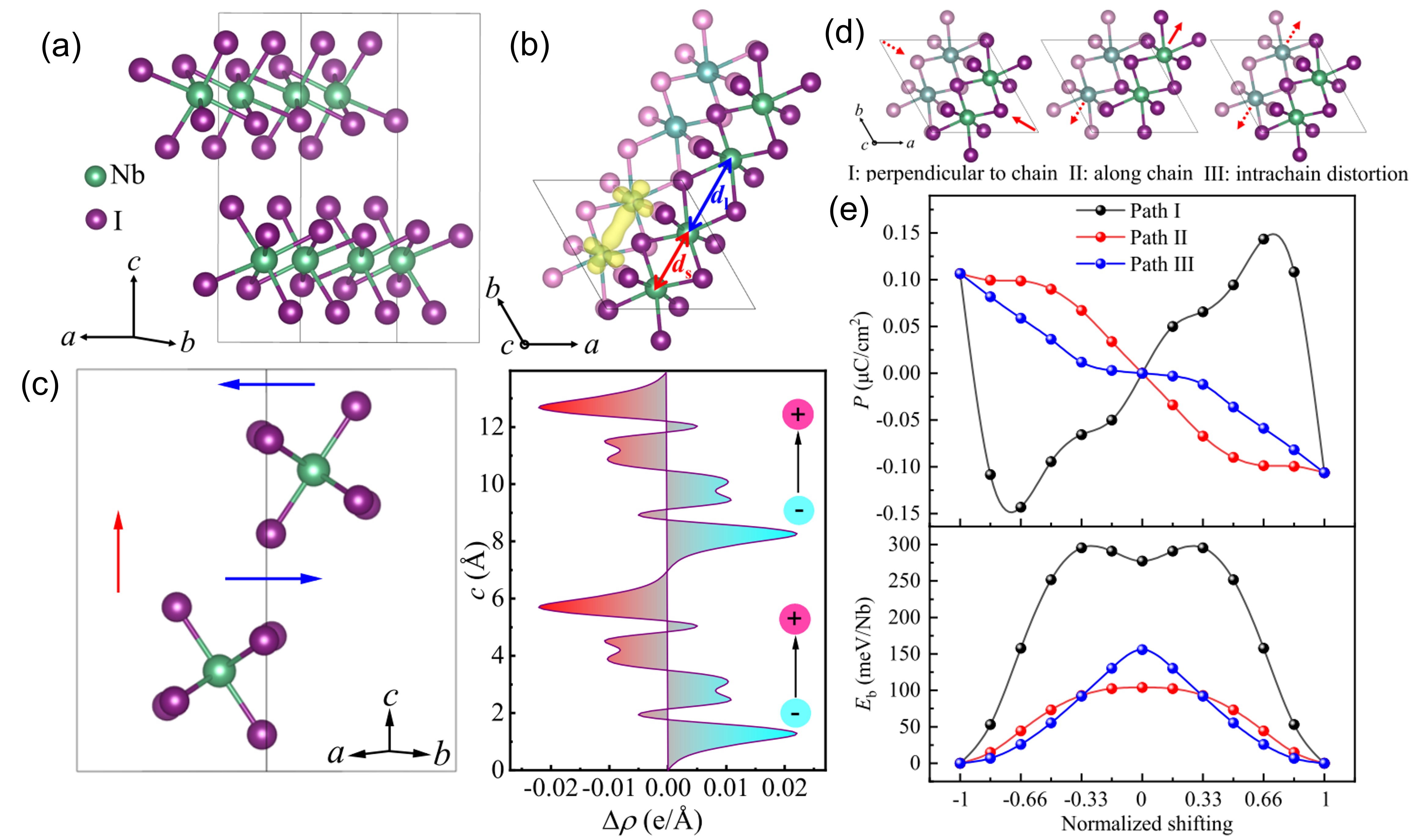}
	\caption{(a) Side view of NbI$_4$ bulk. Each unit cell contains two chains. (b) View of NbI$_4$ from the $c$-axis. The longer $d_l$ and shorter $d_s$ indicate the structural dimerization. (c) Left panel: top view of chains, the bule and red arrows indicate the dipoles. Right panel: The planar-averaged differential electron density [$\Delta\rho=\iint[\rho(+P)-\rho(-P)]dxdy$] along the $c$-axis. The bule and red arrows indicate $p_{ab}$ and $p_c$ respectively. (d) Three possible ferroelectric switching paths in NbI$_4$ bulk. (e) Corresponding $P$ and energy barriers for switching paths. Reprint figure with permission from~\cite{ding2024-PRB}.}
	\label{fig-NbI4}
\end{figure}
There is a obvious advantage of high storage density due to quasi-one-dimensional characteristics, for example, the theoretical upper-limit for ferroelectric memory density can reach $\sim 100$ Tbits/in$^2$~\cite{lin2019-PRM}. Naturally, generalizing the concept of sliding ferroelectricity to quasi-one-dimensional is necessary and promising. Recently, the quasi-one-dimensional ferroelectricity in NbI$_4$ bulk has been reported~\cite{ding2024-PRB}.

There are two NbI$_4$ chains assembled by vdW force in the primitive cell of NbI$_4$ bulk. Within each chain, every Nb ion is caged within the I$_6$ octahedron and the neighbor octahedra are edge-sharing. In addition, the adjacent Nb ions are dimerized as depicted in Fig.~\ref{fig-NbI4}(a,b). The space group of NbI$_4$ bulk is $Cmc2_1$, which belongs to the polar point group $mm2$. The polarization can be calculated to $0.11$ $\mu$C/cm$^2$, whose direction of polarization is in the $c$-axis. Similar to the two-dimensional interlayer sliding ferroelectrics, the special binding mode of isolated two chains can induce the perpendicular components ($p_c$ \& $p_{ab}$). Furthermore, the $p_c$'s are parallel between nearest-neighbor pairs, while $p_{ab}$'s are antiparallel and cancelled between nearest-neighbor pairs along the $c$-axis [Left panel of Fig.~\ref{fig-NbI4}(c)]. The microscopic origin of dipoles is the bias of electron cloud as described in the right panel of Fig.~\ref{fig-NbI4}(c).

Different from the interlayer sliding ferroelectrics, there are more degrees of freedom regarding the sliding modes due to the quasi-one-dimensional characteristic of NbI$_4$. Specifically, there are three paths can reverse the polarization as depicted in Fig.~\ref{fig-NbI4}(d), including sliding along the direction of chain (labeled as I), perpendicular to chain (labeled as II), and intrachain ion displacements (labeled as III). The corresponding polarization and energy barriers for these three switching paths are shown in Fig.~\ref{fig-NbI4}(e). In addition, the negative piezoelectric effect has also been observed in NbI$_4$ system, with $d_{33}$=$-0.42$ pC/N, and the mechanism is similar to that of the interlayer sliding ferroelectricity. Namely the electric dipoles come from interchain couplings, which can be enhanced during the volume shrinking. However, the polarization and negative piezoelectric coefficient are very small, which may limit the practical application. Thus, finding systems with better negative piezoelectric property is one of the research tasks.

\subsection{In-plane negative piezoelectricity in elemental ferroelectrics}
\subsubsection{Theoretical prediction of ferroelectricity in elementary ferroelectric monolayers}
\begin{figure}
	\centering
	\includegraphics[width=\textwidth]{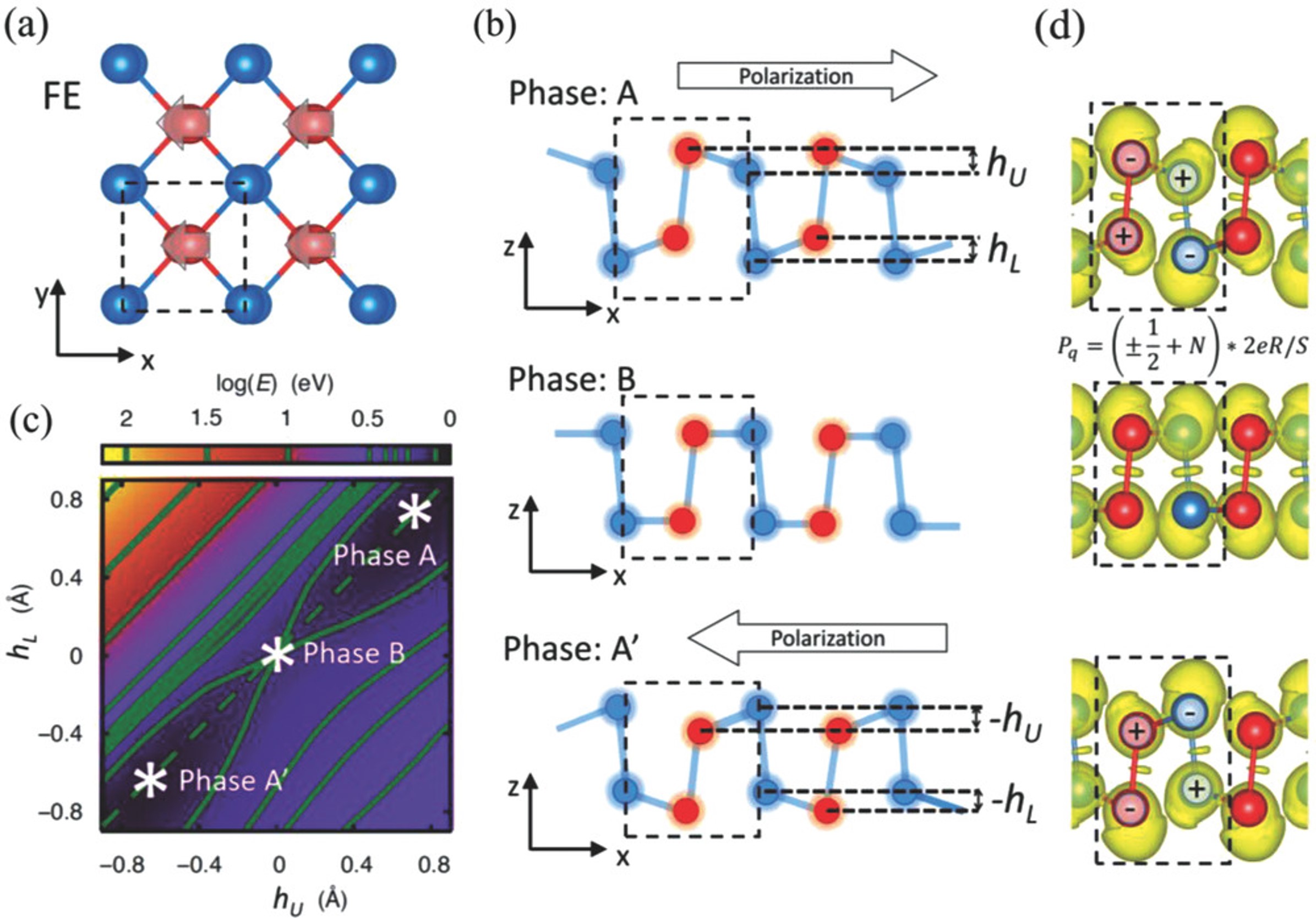}
	\caption{(a) Top view of group-V elemental monolayer. The unit cell is framed by the black dashed rectangle. (b) Side views of ferroelectric states (phases A and A') and paraelectric state (phase B). (c) Free energy contour for As monolayer as the function of buckling heights ($h_U$, $h_L$). (d) The electron localization function for phases A, B, and A'. Reprint figure with permission from~\cite{Xiao2018-AFM}.}
	\label{fig-Bi-FE}
\end{figure}
In recent years, the intrinsic two-dimensional ferroelectricity has been discovered for single-element monolayers by first-principles calculations and experiments, which are group-V elements including As, Sb, and Bi~\cite{Xiao2018-AFM,wang2023-PRB,zhong2023-PRL,Hong2024-PRB}. The structure of elemental monolayers are similar to the buckled black-phosphorene as depicted in Fig~\ref{fig-Bi-FE}. due to the partial $sp^2$ character other than the homogenous tetrahedral $sp^3$ hybridization. The space group of the distorted elemental monolayers is $Pmn2_1$, which is the polar group. And the buckling can be characterized by the two heights labeled $h_U$ and $h_L$. The structure tends to the phosphorene structure ($Pmna$) with centrosymmetry. And there are two stable energy-degenerate polar phases named phase A and A'. The electron localization function shows the distinct localized lone pair in these two polar states. And the polarizations estimated are $46$, $75$, and $151$ pC/m for monolayer As, Sb, and Bi respectively. The polarization can be switched through the paraelectric phase B with a moderate energy barrier. Furthermore, the in-plane polarization of single-element bismuth monolayer has been confirmed by experiment~\cite{gou2023-Nature}.
	
\subsubsection{Large in-plane negative piezoelectricity}

The special buckling distortion induces the ferroelectricity in single-element materials, which can be significantly affected by external strain. The piezoelectric strain coefficients $d_{33}$ of two-dimensional elemental-ferroelectrics can be calculated by the first-principles calculations, which are $-19.2$ and $-25.9$ pC/m for $\alpha$-Sb and $\alpha$-Bi monolayers respectively. The results indicate the large in-plane negative piezoelectricity in these two-dimensional elemental-ferroelectrics~\cite{wang2023-PRB}.
\begin{figure}
	\centering
	\includegraphics[width=\textwidth]{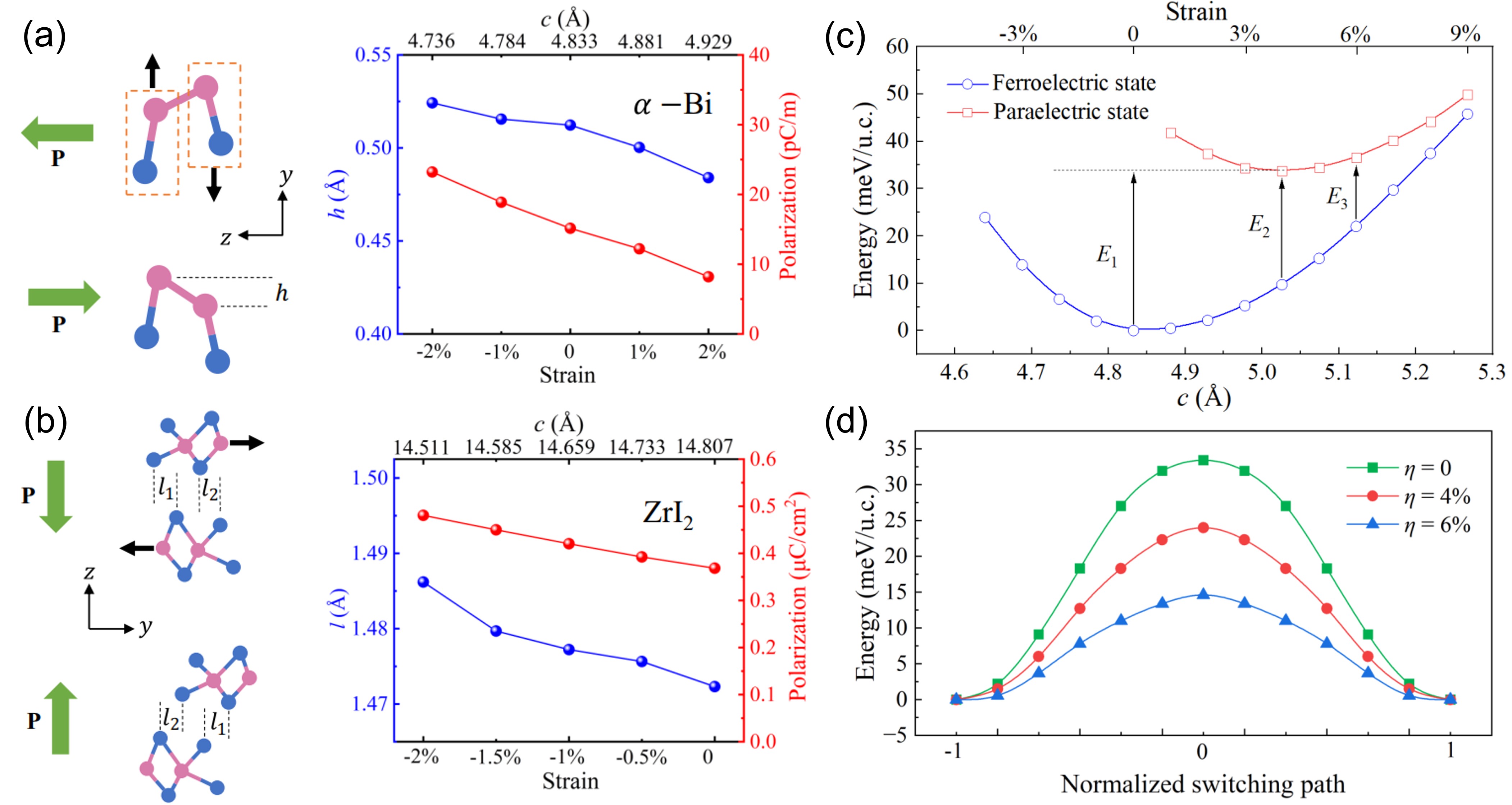}
	\caption{Left panels: the ferroelectric distortion parameters, namely $h$, $l$=$l_1$+$l_2$. $l$ is the sliding distance. The black arrows indicate the directions of atomic movement. Green arrows show the direction of polarization. Right panels: ferroelectric order parameters and polarization as a function of uniaxial strain. (a) Monolayer $\alpha$-Bi. (b) ZrI$_2$ bulk. Reprint figure with permission from~\cite{wang2023-PRB}.}
	\label{fig-Bi-pie}
\end{figure} 

The mechanism of in-plane negative piezoelectricity in these materials may be tracked back to the origin of ferroelectricity. As mentioned before, the ferroelectric phase with buckling distortion breaks the space inversion symmetry and induces charge asymmetry in the puckered monolayer. From another perspective, the buckling is similar to sliding, which can be considered as a kind of intercolumn sliding as shown in Fig.~\ref{fig-Bi-pie}(a,b). The origin of negative piezoelectricity can be revealed qualitatively. The relative sliding between columns induces the polarization perpendicular to the column direction. The neighbor columns become closer with compressive strain along the direction of polar axis, and the polarization enhances with greater ferroelectric displacement.

Furthermore, the piezoelectric stress coefficients $e_{33}$ can be consisted of two terms, namely the clamped-ion part $\bar{e_{33}}$ and the internal-strain part $e'_{33}$. Specifically, $\bar{e_{33}}$ indicates the variety of polarization induced by the distortion of lattice with atomic sites fixed, which indicates the effect of electrons' redistribution and the variety of Born effective charge under uniform strain. And $e'_{33}$ is the piezoelectric reply to the structural optimization that emancipate the internal strain, which means the Born effective charges are fixed. The coefficients of monolayer $\alpha$-Bi are $\bar{e_{33}}=0.6$, $e'_{33}=-5.7$ in unit of $10^{-10}$C/m. The internal-strain is negative with larger value while the clamped-ion is positive with small value. Thus, it is obvious that the negative internal-strain $e'_{33}$ plays the major role in the negative piezoelectricity. This behavior is similar to the vdW layered polar materials BiTe$X$~\cite{kim2019-PRB}, but different from the negative piezoelectricity in the three-dimensional $ABC$ ferroelectric \cite{liu2017-PRL}. Besides, the negative piezoelectric effect caused by the negative clamped-ion term being larger than the internal strain term can exist in monolayer arsenic chalcogenides and monolayer group IV–V $MX_2$ \cite{Gao2020-NL,zhao2021-JMCC,xu2020-APL}.  
	
Similar to WO$_2$$X_2$ monolayer as mentioned above, the shrunk lattice constant in the polar axis is also discovered in the elemental ferroelectrics as an interesting consequence of negative piezoelectricity. As described in Fig.~\ref{fig-Bi-pie}(c), the lattice constant $c$ of paraelectric state is longer than the ferroelectric state, which induces a tensile strain to decrease the energy gain from the ferroelectric displacement. And this can reduce the energy loss when the ferroelectric switching process as shown in the Fig.~\ref{fig-Bi-pie}(d).

\subsubsection{Nonanalyticity of piezoelectric response}

The nonanalyticity of piezoelectric response is a previouly unknown phenomenon, which means the (negative) piezoelectric coefficients of tensile and compressive strains are different~\cite{zhong2023-PRL}. The origin of this unique property will be explained in detail as the following.

Taking the $M_y$ mirror symmetry of the single-element ferroelectrics into consideration, the internal-strain term $e_{11}^{(i)}$ (same as $e_{33}$ due to the different of crystal axes in different papers) can be decomposed to two terms $e_{11,x}^{(i)}$ and $e_{11,z}^{(i)}$. Therefore, the piezoelectric coefficient $e_{11}$ can be wrote as:
\begin{equation}
	e_{11}=e^{(0)}_{11}+e^{(i)}_{11,x}+e^{(i)}_{11,z},
\end{equation}
where 
\begin{equation}
	e^{(i)}_{11,x}=\frac{qa}{\Omega}\sum_{n}Z^{*}_{11}\frac{\partial u_{1}(n)}{\partial\eta},
	e^{(i)}_{11,z}=\frac{qc}{\Omega}\sum_{n}Z^{*}_{13}\frac{\partial u_{3}(n)}{\partial\eta},
\end{equation} 
the electron charge is marked by $q$, $a$ and $c$ are the lattice constants along the $x$ and $z$ axes, $\Omega$ and $Z^{*}_{ij}$ are the area of unit cell and the Born effective charge, respectively. And $u_i$ is the fractional coordinate along $i$ direction.

\begin{figure}
	\centering
	\includegraphics[width=\textwidth]{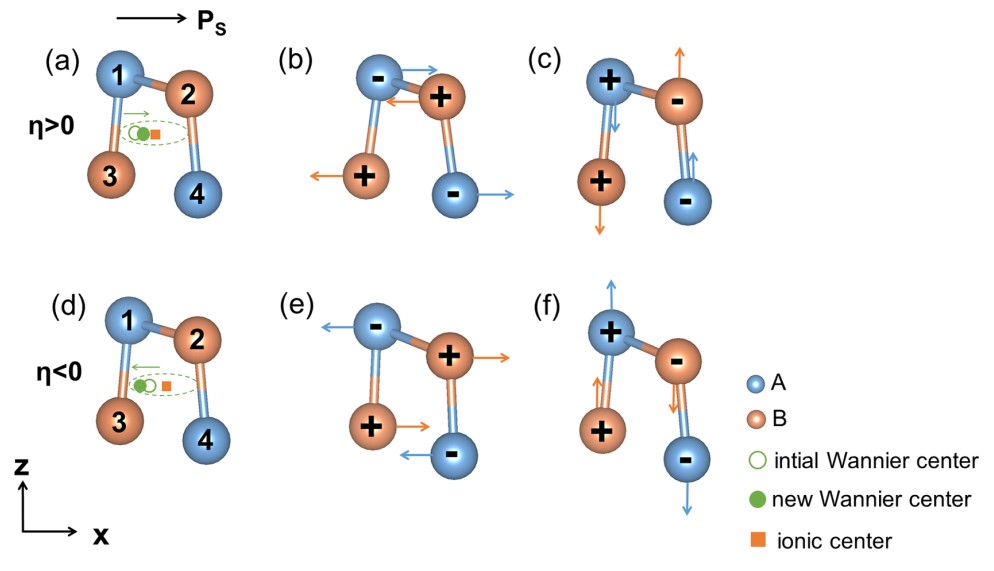}
	\caption{The movements of Wannier center and atomic positions. (a-c) $\eta>0$. (d-f) $\eta<0$. The displacement of Wannier center at the clamped-ion state can be shown by the green dashed oval. The new and initial sites of Wannier center are indicated by the green solid and hollow circle, respectively. The charge center of ions is represented by the orange solid square. The sign of Born effective charges are represented by ``+" and ``-". The movement of anions and cations are indicated by the blue and orange arrows. Reprint figure with permission from~\cite{zhong2023-PRL}.}
	\label{fig-Bi-negative}
\end{figure}  

\begin{figure}
	\centering
	\includegraphics[width=\textwidth]{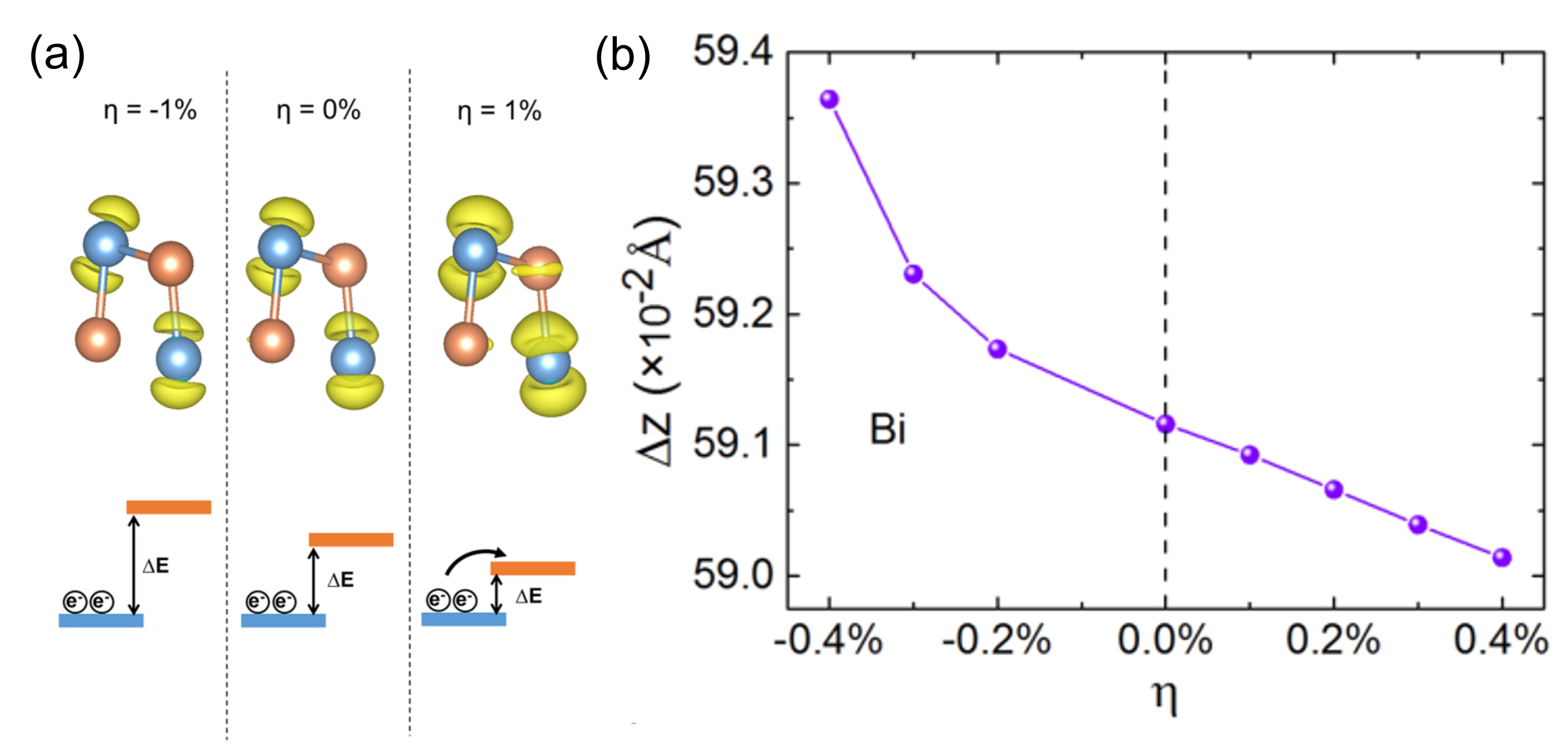}
	\caption{(a) Projected charge densities of $p_z$ orbital for Bi monolayer under different stress conditions. (b) Buckling height $\Delta z$ with the strain $\eta$. Reprint figure with permission from~\cite{zhong2023-PRL}.}
	\label{fig-Bi-unique}
\end{figure}  

Both the clamped-ion term and internal-strain term are negative. First, the negative clamped-ion term indicates ``lag of Wannier center" effect. Specifically, the Wannier center keeps away from the center of positive charge then shifts in the $-x$ direction in the ferroelectric state as depicted by Fig.~\ref{fig-Bi-negative}(a), inducing a polarization along the $+x$ direction ($+P_S$). Imposing a uniform tensile strain $\eta>0$, Wannier center lags behind and has a relative movement in the direction of polarization, which reduces the net polarization and results in negative $e_{11}^{(0)}$. Second, the negative internal-strain terms $e^{(i)}_{11,x}$ and $e^{(i)}_{11,z}$ are related to the special buckled distortion. There is charge transfer in the ferroelectric state [Fig.~\ref{fig-Bi-negative}(b)], The orange and bule sites represent positive and negative charges, respectively. When $\eta>0$, internal-strain allows the shift of ions along the $x$ and $z$ [Fig.~\ref{fig-Bi-negative}(b) and (c)]. Two sublattices move toward each other in the $x$ direction, which decreases the polarization $P_S$. And the ions move toward the paraelectric state in the $z$ direction, which also reduces the polarization.

In Bi monolayer, the calculation indicates that both the clamped-ion term $e^{(0)}_{11}$ and the internal-strain term $e^{(0)}_{11,z}$ are very different under tensile and compressive strains, but $e^{(0)}_{11,x}$ is almost unchanged~\cite{zhong2023-PRL}. This phenomenon can be named as the nonanalyticity of piezoelectric response, which is closely related to the buckling in Bi monolayer. There is a weak hybridization between $6s$ and $6p$ orbitals in Bi monolayer, and $p$ orbitals mainly contribute the bonds between Bi atoms. Due to the Bi monolayer is extended in the $x-y$ plane, the bonds are mainly determined by $p_x$ and $p_y$ orbitals and $p_z$ electrons form lone pairs, which is the driving force of the buckling ($\Delta z$). In addition, the energy difference of $p_z$ levels on the two sublattices acutely affects the distribution of lone-pair electron, which can be regulated by the strain. For the small tensile strain, the energy difference of $p_z$ orbital on two sublattices can be reduced, which induces charge redistribution and decreasing the charge transfer [Fig.~\ref{fig-Bi-unique}(a)]. However, for the compressive strain, there is no obvious charge redistribution. The lone-pair electrons also affect the height ($\Delta z$). Fig.~\ref{fig-Bi-unique}(b) indicates $\Delta z$ under compressive strain changes more rapidly with $\eta$, which can explain the nonanalyticity of piezoelectric response. 

\section{Summary and outlook}
Negative piezoelectric systems not only have common applications of piezoelectric materials, but also have unique and promising potential applications in devices used for electromechanical systems. Firstly, piezoelectric materials can produce an electrical voltage in response to an external mechanical force, which means energy conversion between mechanical energy and the electrical energy can be achieved. Piezoelectric nanogenerators are the main choice to harvest mechanical vibration energy from the surrounding environment, which has been received great attention due to the higher energy density and conversion efficiency \cite{Liu2018-APR}. PVDF as a typical negative piezoelectric material has been widely used in piezoelectric nanogenerator devices \cite{Chang2010-NL} including piezo-photodetectors \cite{Si2020-Nanoscale,Didhiti2021-Nanoscale}, self-powered sensors \cite{Li2021-Nanoscale}, piezoelectric fibers \cite{Sarah2022-ACS-AMI}, and triboelectric textile sensors \cite{Lee2021-CPBE}. In addition, the negative piezoelectricity can be an important factor for output performance of PVDF-based piezoelectric hybrid nanogenerators \cite{Guo2022-ACS-AMI}. Secondly, the negative piezoelectric effect has exclusive and promising potential application owing to the difference between the response of negative piezoelectric materials to external electric fields and that of positive piezoelectric materials. As depicted in Fig. 1 (b), a heterostructure composed by ultrathin positive piezoelectric and negative piezoelectric layers can be designed, when an external electric field is applied to this heterostructure, then the lattice of normal piezoelectric layer expanses but that of negative piezoelectric layer shrinkages. Thus, very large bending in materials can be achieved by constructing above heterojunctions \cite{wang2023-PRB}.
 
In this topical review, the concept of negative piezoelectric effect and the research progress of two-dimensional ferroelectrics were introduced. Furthermore, the negative piezoelectricity in quasi-two/one-dimensional ferroelectrics reported in recent works were reviewed case by case. The mechanism of negative piezoelectricity is highly different, which is closely related to the origin of ferroelectricity. It can be roughly categorized into the following four types. First, in CIPS system, the soft vdW layer is responsible for the volume shrinking upon pressure while the electric dipoles are from non vdW layer. Second, the negative piezoelectricity in two-dimensional $M$O$_2$$X_2$ is due to the noncollinearity of local dipoles, which results in the orthogonal ferroelectric and antiferroelectric axes. Third, the origin of negative piezoelectricity in interlayer/quasi-one-dimensional sliding ferroelectrics is the joint contribution of dipole moment increase and volume decrease. Fourth, the large in-plane negative piezoelectricity in monolayer Bi is highly related to the buckling structure. 

Research on piezoelectric property in low-dimensional ferroelectrics is in the ascendant, and experience can be learned from perovskite systems. Previous studies indicate the high piezoelectricity in traditional ferroelectrics including Pb(Zr,Ti)O$_3$ (PZT) \cite{Service1997-science}, Pb(Mg$_{1/3}$Nb$_{2/3}$O$_3$)-PbTiO$_3$ (PMN-PT) \cite{Park1997-JAP} and Sm-doped PMN-PT system \cite{Li2019-science, Li2018-NM, Kumar2020-JPDAP} benefits from the morphotropic phase boundary (MPB) and nanoscale structural heterogeneity \cite{Li2019-science}. Although some works have focused on MPB in the low-dimensional ferroelectrics \cite{song2021-PRB,Deng2024-2D-Materials,wang2024-NL}, achieving phase coexistence and designing possible polarization rotation is still a challenge for vdW ferroelectrics. Whether a giant negative piezoelectric effect in low-dimensional ferroelectrics can be obtained through MPB remains to be further studied. In addition, the previous work has been reported that a highly polarizable ionic-covalent matrix benefits the giant piezoelectricity in a non-MPB ferroelectric xBi(Ni$_{0.5}$Hf$_{0.5})$O$_3$-(1-x)PbTiO$_3$, which has been revealed by the techniques like neutron pair distribution functions and complementary Raman scattering measurements \cite{Datta2021-PRB-L}. Besides, the domain switching and lattice strain are important processes contributing to the large electromechanical response in perovskite-based piezoelectrics, and the coupling between them has also been reported in previous works \cite{Hall2004-JAP,Ranjan2018-PRB}. These research experiences can be used for reference in the study of two-dimensional piezoelectric materials.

With the exception of CIPS and two-dimensional ferroelectric metal distorted MoS$_2$ \cite{Lu2023-ACS-ML} reported by recent experiments, these studies about the negative piezoelectric effect mainly performed by first-principles calculations. In other words, there are still very few negative piezoelectrics that have been experimentally verified, which requires further research and exploration. In addition, the mechanisms of negative piezoelectric effect should be more diverse with the increasing number of discoveries of low-dimensional ferroelectric materials. Finally, devices design related to the negative piezoelectric effect of low-dimensional ferroelectrics is lacking, which will open a new scientific era.

\section*{Acknowledgment}
We thank Prof. Junling Wang, Prof. Lu You, Dr. Ling-Fang Lin, Dr. Yang Zhang, Haoshen Ye, and Ziwen Wang for helpful discussions. This work was supported by the National Natural Science Foundation of China (Grant Nos. 12325401, 12274069) and Jiangsu Funding Program for Excellent Postdoctoral Talent under Grant Number 2024ZB001. 

\bibliography{IOP}
\bibliographystyle{iopart-num}		
\end{document}